\documentclass[aip,cha,amsmath,amssymb,reprint,
]{revtex4-1}
\usepackage[version=3]{mhchem}
\usepackage{graphicx}
\usepackage{dcolumn}
\usepackage{bm}
\usepackage[colorlinks=true,linkcolor=blue]{hyperref}

\begin{document}
\preprint{AIP/123-QED} 

\title[Reconstructing the free-energy]{Reconstructing the free-energy landscape\\ associated to molecular motors processivity}

\author{J. L\'opez-Alamilla}
\affiliation{Facultad de Ciencias, Universidad Nacional Aut\'onoma de M\'exico, Circuito exterior de Ciudad Universitaria, M\'exico Distrito Federal 04530, M\'exico}
\author{I. Santama\'ia-Holek}%
 \email{isholek@gmail.com}
\affiliation{UMJ-Facultad de Ciencias, Universidad Nacional Aut\'onoma de M\'exico, Campus Juriquilla, Quer\'etaro 76230,
M\'exico}

\date{\today}

\begin{abstract}
We propose a biochemical model providing the kinetic and energetic descriptions of the processivity dynamics of kinesin and 
dinein molecular motors. Our approach is a modified version of a well known model describing kinesin dynamics and considers the presence of a competitive 
inhibition reaction by ADP. We first first reconstruct a continuous free-energy landscape of the cycle catalyst process that allows us to calculate the 
number of steps given by a single molecular motor. Then, we calculate an analytical expression associated to the translational velocity and the 
stopping time of the molecular motor in terms of  time and ATP concentration. An energetic interpretation of motor processivity is discussed in quantitative 
form by using experimental data. We also predict a time duration of collective processes that agrees with experimental reports.
\end{abstract}

\maketitle
\section{Introduction}\label{sec:I}

Molecular motors are ubiquitous entities in living cells that participate in controlling and synchronizing 
many cellular transport processes and are involved in different metabolic pathways~\cite{Vale}. This essential role makes the study of their 
biochemical kinetics of relevant interest in biology and medicine \cite{lehninger}. In addition, the cyclic operation of these motors and 
its possible control may be of help for developing future externally controllable nano-machines ~\cite{Peterman,Switch}. 

Many experimental and theoretical studies have been devoted to understand the detailed biochemical reactions associated to the motion of kinesins and other molecular motors, see, for instance, Refs. 
\cite{Peterman,peterman2007,Visscher1999,Visscher2000,Progresion-Kinesin,ADP,ADP-production-1,ADP-production-2,nature-Inhibition}. 
One of the aims of these studies is to characterize protein conformational changes in order to understand the sequence
of reactions accompanying motor translation and determining its energetic dependence \cite{Visscher1999,Visscher2000,Progresion-Kinesin,svoboda}. To this end,  
single-molecule fluorescence spectroscopy and polarization microscopy techniques have been used in order to better discriminate individual motor activities
due to the large heterogeneity in motor behavior \cite{Peterman}. These studies allow an accurate determination of step sizes, rates and pausing properties 
of different motor dynamics that can be used to propose more specific theoretical models \cite{Visscher1999,ADP,ADP-production-1,ADP-production-2,nature-Inhibition,Svoboda1993,Hanggi,Hanggi1}.  Often, these models follow two different approaches, one describing the collective 
behavior and properties by viewing the motors as catalytic agents of a chemical reaction \cite{Visscher1999,Svoboda1993} whereas the second one 
focuses on the individual description that analyzes the specific conformational changes occurring during motility \cite{ADP,ADP-production-1,ADP-production-2,nature-Inhibition}. 
From a more physical point of view, several models taking into account the stochastic and thermodynamic nature of the translational dynamics of molecular motors  
were proposed and used to describe some general features of molecular motor activity~\cite{Hanggi,Hanggi1,Reimann,Parrondo,Rubi,Rubi1,ernesto}. 
The essential ingredient to perform the description in these approaches is to provide a model for the energy landscape of the cyclic operation of motors
\cite{Hanggi,Hanggi1,Reimann,Rubi1,ernesto}, however these models are barely related to the chemical reactions that drive the process.

Despite of the particular mechanisms determining the rate at which each catalyst cycle occur, some general
characteristics can be used in order to reconstruct the energy landscape of the entire process, 
that is, they allow for determining the processivity of the molecular motor under consideration. Understanding the
kinetic and energetic properties associated to the processivity of molecular motors is of central importance when 
connecting the specific dynamics to the biological function \cite{lehninger,Alberts}. This is because cellular transport 
processes are usually driven by more than one motor, therefore requiring a high degree of cooperativity.

As proteins, enzymes and molecular complexes, the processes developed by molecular motors use chemical energy 
stored into molecules such as ATP or GTP, produced by the mitochondrial system of the cell \cite{lehninger}.  However, 
many studies are performed  \emph{in vitro}  with well controlled ATP concentrations and somehow simplified 
conditions concerning the viscoelastic properties of the surroundings. This allows for a better analysis of the detailed 
dynamics of the motor but may hide some aspects of its performance \emph{in vivo}, such as the transport of proteins, 
RNA, vesicles and even organelles \cite{embjo} that may be related to, for instance, several exocytosis-endocytosis processes \cite{steinhardt,epithelial,Allosteric,neuro}. 

In this work, we analyze the thermodynamics and chemical kinetics of two theoretical models describing motor processivity \cite{Visscher1999,ADP,hanckock}, 
their time activity and the associated energy consumption. These two hand-over-hand models were first developed to describe the 
particular problem of intracellular transport via kinesin along microtubules and are based on the well known evidence that the energy used for the motion 
comes from ATP hydrolysis~\cite{Visscher1999,Visscher2000,ADP,Progresion-Kinesin}.  We first simplify the general schemes proposed in the literature
by assuming that the three initial steps of the reaction sequence are slower than the last ones, and therefore determine the time step of the overall reaction. 
In a second model, we also assume that the ADP produced during the motion of the molecular motor plays the role of
a competitive inhibitor \cite{ADP,ADP-production-1,ADP-production-2,nature-Inhibition}. In both cases, we use the concept of the degree of reaction 
\cite{Prigogine} and follow the rules of  
thermodynamics to reconstruct the corresponding Gibbs free-energy landscape for a single cycle and motor. In this way, the free-energy we obtain 
is a quantitative representation of an enzyme catalytic reaction in which protein conformation fluctuations in the presence of the 
substrate are taken into account, that is, it constitutes a model of an enzyme complementary to the transition state and not to the substrate \cite{lehninger}. 

Then, using Fourier analysis we are able to reconstruct the complete energy landscape of the whole catalytic reaction and the translational velocity
of the motors through a well established procedure for the analysis of the enzymatic reaction scheme. By its nature, this second information is 
important because complements that of the free-energy by giving a collective notion associated to motors activity. In the case of the process 
with inhibition, we also determine the time course of both, the free energy and the translational velocity. We analyze the effect
of different ATP and ADP levels on this velocity. In addition, we also give a simple expression for the stopping time of a collectivity of molecular motors, 
an important quantity characterizing the finite processivity of the motors when participating in metabolic tasks. These results allow us to predict the number
of steps that a single motor may perform before stopping, and also the average traveled distances and the associated ATP consumptions of a 
collectivity of motors giving the initial values of the different parameters entering in the description. The effect of the load is also taken into account. 

\begin{figure*}[]
\begin{center}
\includegraphics[width=400pt]{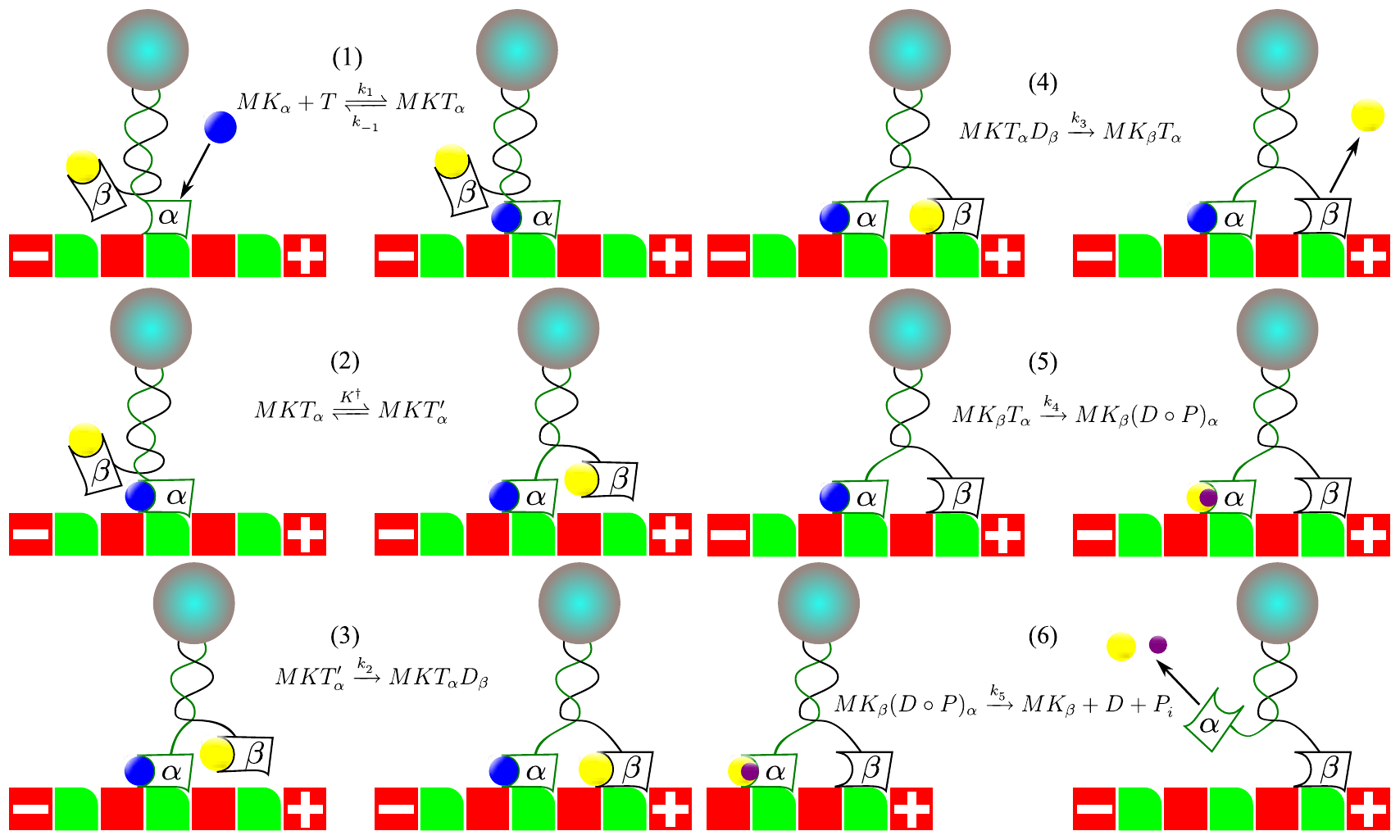}
\caption{Schematic representation of Eqs. (1)-(6). Blue molecules represents ATP, yellow molecules represents ADP, purple molecules inorganic phosphate.}
\label{fig:1}
\end{center}
\end{figure*}

The article is organized in five sections. In Section~\ref{sec:II} we present the essential biochemical reaction model that we will modify accordingly with the aim
to determine, in Section 3, the continuous free-energy landscape of the catalytic reaction and the average translational velocity of a kinesin motor. Section~\ref{sec:IV} is then devoted to analyze the inhibitory effect of ADP as the main energetic mechanism determining the finite time of motor processivity. The
time dependence of the translational velocity and the stopping time are also calculated and discussed using values for the experimental parameters 
taken from literature.
\section{The hand-over-hand mechanism}\label{sec:II}

The two models accounting for the transport process via kinesin activity that we will discuss in the following sections are simplifications based on the original biochemical reaction scheme proposed in Ref.~\cite{Visscher2000} and based on detailed experimental studies cite{Visscher1999,ADP,svoboda,hanckock}. The first reaction of a sequence of six (as shown 
schematically in Figure 1) accounts for the capture of an ATP molecule ($T$) by the microtubule-kinesin complex $MK$ in order to produce the enzyme-substrate complex $MKT_\alpha$
\begin{equation}\label{rec:1}\displaystyle
MK_\alpha^{}+T\ce{<=>[k_1][k_{-1}]}MKT_\alpha^{}\,,
\end{equation}
where the subindex $\alpha$ indicates that the corresponding kinesin head $K_\alpha$ is attached to the microtubule, whereas the  second head $\beta$ is free. According to this scheme, the complex $MK$ plays the role of an enzyme that acts over substrate $T$ through a catalytic reaction. This means that free kinesins cannot consume ATP periodically in time unless they are attached to a microtubule~\cite{Visscher2000}. The complete reaction sequence associated to a single step is given through Eqs. (\ref{rec:2})-(\ref{rec:6})
\begin{equation}\label{rec:2}\displaystyle
MKT_\alpha^{}\ce{<=>[K^{\dag}][]}MKT_\alpha^{\prime}\,,
\end{equation} 

\begin{equation}\label{rec:3}\displaystyle
MKT_\alpha^{\prime}\ce{->[k_2][]}MKT_{\alpha}D_{\beta}\,,
\end{equation}

\begin{equation}\label{rec:4}\displaystyle
MKT_{\alpha}D_\beta\ce{->[k_3][]}MK_{\beta}T_{\alpha}\,,
\end{equation}

\begin{equation}\label{rec:5}\displaystyle
MK_{\beta}T_{\alpha}\ce{->[k_4][]}MK_{\beta}(D\circ{P})_\alpha\,,
\end{equation}

\begin{equation}\label{rec:6}\displaystyle
MK_{\beta}(D\circ{P})_\alpha\ce{->[k_5][]}MK_{\beta}+D+P_i\,.
\end{equation}

In accordance with experimental observations~\cite{Visscher1999,Progresion-Kinesin,hanckock}, the last reaction reflects the fact that the hydrolysis of ATP at the active site $\alpha$ of the kinesin produces enough energy in 
order to liberate the corresponding head. After this reaction occurs, the cycle is completed and an initial state $MK_\beta$ is recovered one step forward from the initial position and with lower free-energy ref.~\cite{Visscher2000}.  
The number of repetitions of this cycle that the molecular motor is capable to perform determines its processivity.

Here, it is important to stress that reaction (\ref{rec:6}) produces an ADP molecule and an 
inorganic phosphate $P_i$ that can also react with the corresponding active head of the kinesin \cite{ADP}. As a 
consequence of this, a complex inhibition scheme controlling motor processivity may arise \cite{ADP,Yajima,Romberg,premature}. Since during the evolution in time of the process ADP and $P_i$ concentrations grow, 
then the probability of occurrence of an inhibition event increases. This process and its influence on motor processivity will be analyzed in detail in the following sections. 
\section{Gibbs free-energy landscape reconstruction}\label{sec:III}
In order to reconstruct the free-energy landscape associated to the motion of the molecular motor as modeled by Eqs. (\ref{rec:1})-(\ref{rec:6}), we will first assume that reactions (\ref{rec:4})-(\ref{rec:5}) are much faster than reactions (\ref{rec:1})-(\ref{rec:3}) ~\cite{Visscher1999,Visscher2000}, and therefore its contribution to the total time of the global reaction can be neglected. As previously mentioned, reactions (\ref{rec:1})-(\ref{rec:3}) are associated to the formation of the active enzyme $MK$ by fixation of kinesin $\alpha$-head to the microtubule in the presence of an ATP molecule, reaction (\ref{rec:1}). This process produces the formation of the enzyme-substrate complex $MKT_\alpha$  that may fluctuate allowing a conformational change of the dimer constituting the motor stalk and producing the secondary enzyme-substrate complex $MKT_\alpha^\prime$, reaction (\ref{rec:2}). The secondary complex has the appropriate structural conformation in order to attach the $\beta$-head to the microtubule
and rapidly hydrolyze the ATP of site $\alpha$ in order to promote the translational motion, reactions (\ref{rec:3})-(\ref{rec:6}). 
According to this model, the isomerization reaction (\ref{rec:2}) is the key ingredient in order to reconstruct the Gibbs 
free energy landscape and implies the presence of a small energy barrier that can be thermally overcome.

If we assume that the reactions taking place during the motion of the molecular motor are not far from equilibrium, we may use non-equilibrium thermodynamics 
to analyze the chemical kinetics of the process \cite{Prigogine}. For each elementary reaction, the differential variation of  the Gibbs free-energy in terms of the 
degree of reaction $\xi$ can be written in the form 
\begin{equation}\label{dG:0}\displaystyle
dG_j=-\sum_i\nu_{ij}\mu_{ij}d\xi_j\,,
\end{equation}
where $\nu_{ij}$ and $\mu_{ij}$ are  the stoichiometric coefficient and the chemical potential of the $i$-th species in the $j$-th reaction. The relation 
between the mass fraction $n_i$ and the corresponding degree of reaction $\xi_j$ is: $dn_i=-\nu_{ij}d\xi_j$,~\cite{Prigogine}. Thus, assuming that 
the system is sufficiently diluted we have
\begin{equation}\label{dG:1}\displaystyle
dG_j=-RT\sum_i\nu_{ij}\ln\left|\frac{n_i}{n_i^{eq}}\right|d\xi_j\,,
\end{equation}
where we used the expression: $\mu_i=RT\ln|n_i/n_i^{eq}|$, with  $n_i$ the molar fraction of the $i$-th chemical specie and $n_i^{eq}$ its 
equilibrium value. 

In this form the change of Gibbs free-energy of reaction (\ref{rec:1}) can be written in the form 
\begin{equation}\label{dG:2}\displaystyle
dG=-RT\ln \left|\frac{n_{MK}^{\nu_{MK}}}{n_{MK_{eq}}^{\nu_{MK}}}\cdot\frac{n_{T}^{\nu_{T}}}{n_{T_{eq}}^{\nu_{T}}}
\cdot\frac{n_{MKT}^{\nu_{MKT}}}{n_{MKT_{eq}}^{\nu_{MKT}}}\right|d\xi \,.
\end{equation}
Here, the relation between $\xi$ and the corresponding molar fractions is established in Table~\ref{tab:1}  for a time $\tau$ elapsed after the 
chemical reaction started.
\begin{table}[!ht]
\begin{tabular}{lcccc}
\hline\hline
\multicolumn{5}{l}{Time \hfill Molar fractions\hfill \ \ }\\[1pt]
\hline\hline\\[1pt]
$\ t=0$&$\ n_{MK}^o$&$\ n_{T}^o$&$\ n_{MKT}^o$&$\ n_{MKT^\prime}^o$\\[3mm]
$\ t=\tau$&$\ n_{MK}^o-\xi$&$\ n_{T}^o-\xi$&$\ n_{MKT}^o+\xi$&$\ n_{MKT^\prime}^o-\xi$
\end{tabular}
\caption{Stoichiometric ratio of chemical species. From this table it follows that for reaction (\ref{rec:1}): 
$\nu_{MK}=1$, $\nu_{T}=1$ and $\nu_{MKT}=-1$. The last column takes into account the fact that the 
equilibrium reaction (\ref{rec:2}) also produces $MKT$, and therefore $\nu_{MKT^\prime}=1$. 
The superscript $^o$ indicates the initial values of the mass fraction the corresponding species.}
\label{tab:1}
\end{table}

Now, taking into account reaction (\ref{rec:2}),  Eq. (\ref{dG:2}) can be rewritten by using the fact that the complexes 
$MKT$ and $MKT^\prime$ are equilibrated according to reaction (\ref{rec:2}), see the Appendix~\ref{app:A} 
for details. As a consequence of this, they obey the relation:  $K^\dag=n_{MKT^\prime}/n_{MKT}$, where $K^\dag=K_oe^{-f\delta/k_BT}$ is the equilibrium constant corrected by the factor $e^{-f\delta/k_BT}$ which takes into account the effect of the load $f$ (weight) imposed by the cargo on the motor  and $\delta$ corresponds to the distance travelled by the cargo during the isomerization \cite{Visscher2000}, typically one half of the characteristic step distance of $8\,nm$. 
Thus, for cargo dependent motions we have in turn 
\begin{equation}\label{eq-rel-1}\displaystyle
n_{MKT}^{eq}=\frac{n_{MKT^\prime}^{eq}}{K_oe^{-f\delta/k_BT}}\,.
\end{equation}

Substituting now the previous relation into (\ref{dG:2}), using Eq. (\ref{eq:A8}) obtained in the appendix and performing a Taylor expansion 
of the logarithm: $\ln(x)\approx x-1$ for $x\sim1$ (notice that the condition $x=n_i/n_i^{eq}\sim1$ is always obeyed for reactions not far from equilibrium) 
we finally obtain the following expression for the effective Gibbs free-energy change of reactions (\ref{rec:1}) and (\ref{rec:2})
\begin{equation}\label{dG:3}\begin{aligned}\displaystyle
\frac{dG}{RT}=-\left[(n^o_{MK}-\xi)(n^o_{T}-\xi)(n^o_{MKT^\prime}-\xi)\right.\,\,\,\,\,\,\,\,\,\,\\
\,\,\,\,\left.-(n^{eq}_{MK}n^{eq}_{T}n^{eq}_{MKT^\prime})+f\delta/k_BT\right]d\xi\,.
\end{aligned}
\end{equation}

An integration over $\xi$ yields the following expression  for the Gibbs free-energy change $\Delta G(\xi)$
\begin{equation}\begin{aligned}\label{G:1}
\frac{\Delta G(\xi)}{RT}=A\xi^4+B\xi^3+C\xi^2+\left(D-D_{eq}\right)\xi\,\,\,\,\,\,\,\\
\,\,\,\,\,+f\delta/k_BT\xi\,,
\end{aligned}
\end{equation} 
where we have defined the constants
\begin{equation}\begin{aligned}\label{ctes}\displaystyle
A=\displaystyle\frac{1}{4},\,\,\,\,\,\, B=\displaystyle\frac{n_{MK}^{o}+n_{T}^{o}+n_{MKT^\prime}^{o}}{3}\,,\\
C=\frac{1}{2}\left(n^{o}_{MK}n^{o}_{T}+n^{o}_{MK}n^{o}_{MKT^\prime}+n^{o}_{T}n^{o}_{MKT^\prime}\right)\,,\\
D=n_{MK}^{o}n_{T}^{o}n_{MKT^\prime}^{o}\,, D_{eq}=n_{MK}^{eq}n_{T}^{eq}n_{MKT^\prime}^{eq}\,.
\end{aligned}\end{equation}
It is important to emphasize here that coefficients $B$, $C$ and $D$ depend upon the initial concentrations of enzyme $MK$, ATP and $MKT^\prime$ 
whereas $D_{eq}$ depends upon the equilibrium concentration values of these species.
\begin{figure}[tbp]
\begin{center}
\includegraphics[height=200pt,width=185pt]{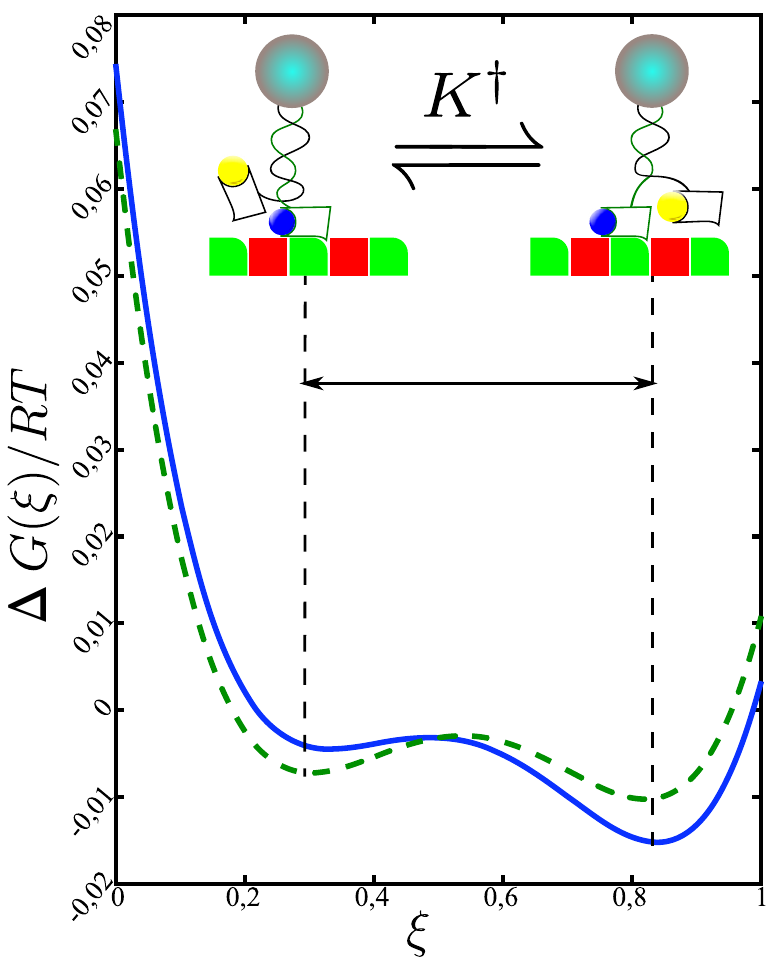}
\caption{The normalized free energy difference $\Delta G$ in terms of the reaction coordinate $\xi$. The blue solid line corresponds to the case without load (cargo weight) whereas the green dashed line corresponds to the case when a cargo of $f=0.02$pN is applied to the molecular motor. The differences of the free energy between states depend on the initial concentrations of enzyme $MK$, ATP and $MKT^\prime$ whereas the global tilting of the potential depends on the equilibrium conditions incorporated by $D_{eq}$. The cargo weight reduces the free energy difference between the isomeric states $MKT$ and $MKT^\prime$, thus decreasing the average translational velocity of the motor. The initial values used are: $\ n_{MKT^\prime}^{o}=~0.02$,$\ n_{T}^{o}=~0.78$ and$\ n_{MK}^{o}=0.2$. The values of the rate constants are given in Fig. 6.}
\label{fig:2}
\end{center}
\end{figure}

Equations (\ref{G:1}) and (\ref{ctes}) constitute the basis for reconstructing the free-energy 
landscape of the catalytic reaction and therefore are useful to explain the processivity of a single molecular motor. Figure~\ref{fig:2} shows the normalized 
Gibbs free energy (\ref{G:1}) in terms of the degree of reaction $\xi$ for values of the parameters taken from 
experiments~\cite{ADP,ADP-production-1,ADP-production-2,nature-Inhibition}. The free-energy difference is tilted and is asymmetric because the left maximum at ($\xi=0$) is higher than the right one ($\xi=1$). In addition, the $MKT^\prime$ state is energetically more favorable than the $MKT$ one, fact that promotes the motion of the motor. In the following sections, we will show that the present scheme for the catalytic reaction can be generalized by considering competitive inhibition by the final products of the sequence of reactions (the production of ADP and $P_i$) and how this leads to invert the above situation, that is, when due to the presence of inhibitors the $MKT$ state becomes energetically more favorable than the $MKT^\prime$ one, which results in the end of the catalytic reaction. In physical terms, these asymmetries shown by the potential imply that the catalytic free-energy landscape we deduced from the biochemical reaction scheme is a type of potential called \lq\lq{tilted ratchet}\rq\rq potential~\cite{Hanggi,Hanggi1,Reimann,Rubi,ernesto}. Thus, the biochemical model proposed here gives an experimental basis to these type of potentials heuristically proposed in the literature.

In Figure~\ref{fig:2}, the local maximum at $\xi\sim0.54$ corresponds to the free energy barrier associated to the isomerization 
reaction (\ref{rec:2}), also represented in the schematic view of the motor on the top of the figure. The corresponding activation energy  
will be denoted by $G^\dag$ and the difference between isomeric states $MKT$ and $MKT^\prime$ will be denoted by $\Delta{G^\dag}$
(see also Figure \ref{fig:2}). 
The solid blue line corresponds to the motion of the kinesin with a cargo of negligible weight. The green dashed line shows that the tilting of the potential decreases 
when the weight of the cargo is increased. The values of the parameters used to represent $\Delta G$ were taken from 
Refs.~\cite{Visscher2000,ADP}.
\subsection{Processivity: The single motor free-energy landscape}\label{sec:IIIA}

The free-energy difference (\ref{G:1}) only applies for describing the isomerization in a single step, and not the overall process associated to the 
displacement of the motor. Thus, in order to model the whole catalytic process, first we have to consider that it constitutes a sequence 
of single steps, one after one. 
\begin{figure}[!ht]
\begin{center}
\includegraphics[width=240pt]{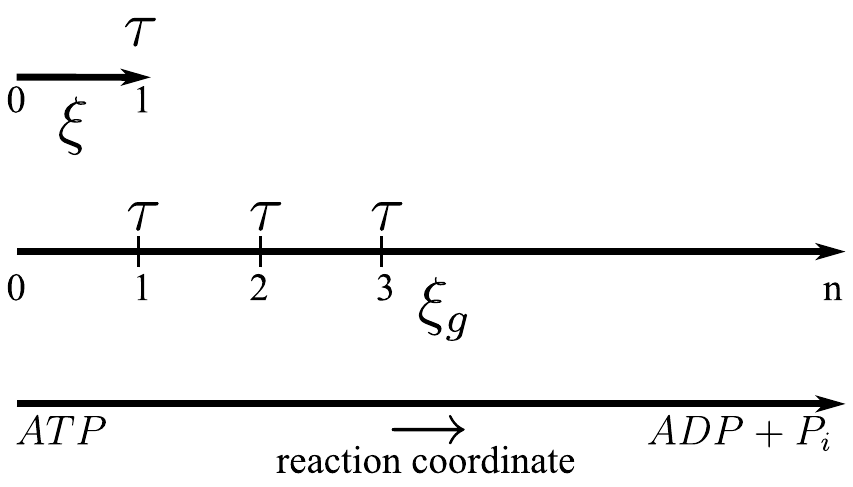}
\caption{Graphical interpretation of the relation between the degree of advance of a single reaction $\xi$, the time duration of each reaction $\tau$ and the global degree of reaction $\xi_g$ corresponding to the catalytic reaction. Each step is characterized by the hydrolysis of one molecule of ATP.}
\label{fig:9}
\end{center}
\end{figure}

Following this idea, the catalyst landscape (cyclic-steps landscape) can be viewed as a periodic potential whose elemental cycle 
is described by Eq. (\ref{G:1}). In order to perform the catalyst landscape reconstruction, we shall generate a periodic potential 
starting from Eq. (\ref{G:1}) and performing a Fourier series expansion considering only the terms depending on the {\em initial values} 
of  the molar fractions $A$, $B$, $C$ and $D$, see Eq. (\ref{ctes}). The other two terms related to $D_{eq}$ and $f\delta/K_BT$ are 
exclusively determined by the equilibrium conditions of the problem and of the load $f$ applied by the cargo.

Hence, the catalyst energy landscape can be expressed in terms of the global degree of reaction $\xi_g$
that goes from $0$ to $\infty$ (as the time variable $\tau$) and which is related with the reaction coordinate $\xi$ as it is shown in 
Figure~\ref{fig:9}.  Each cycle (step) has a duration of $\tau \sim 10\,ms$ and the first step goes from $0$ to $\tau$ whereas the second 
step goes from $\tau$ to $2\tau$, {\em etcetera}. In the present case, the lack of inhibition by the product of the catalytic reaction allows 
to assume that the reaction occurs with constant initial values of the mass fractions $n_{MK}^o$, $n_T^o$ and $n_{MKT^\prime}^o$ 
as illustrated in Table \ref{tab:1}. 
\begin{figure}[!htb]
\begin{center}
\includegraphics[width=230pt]{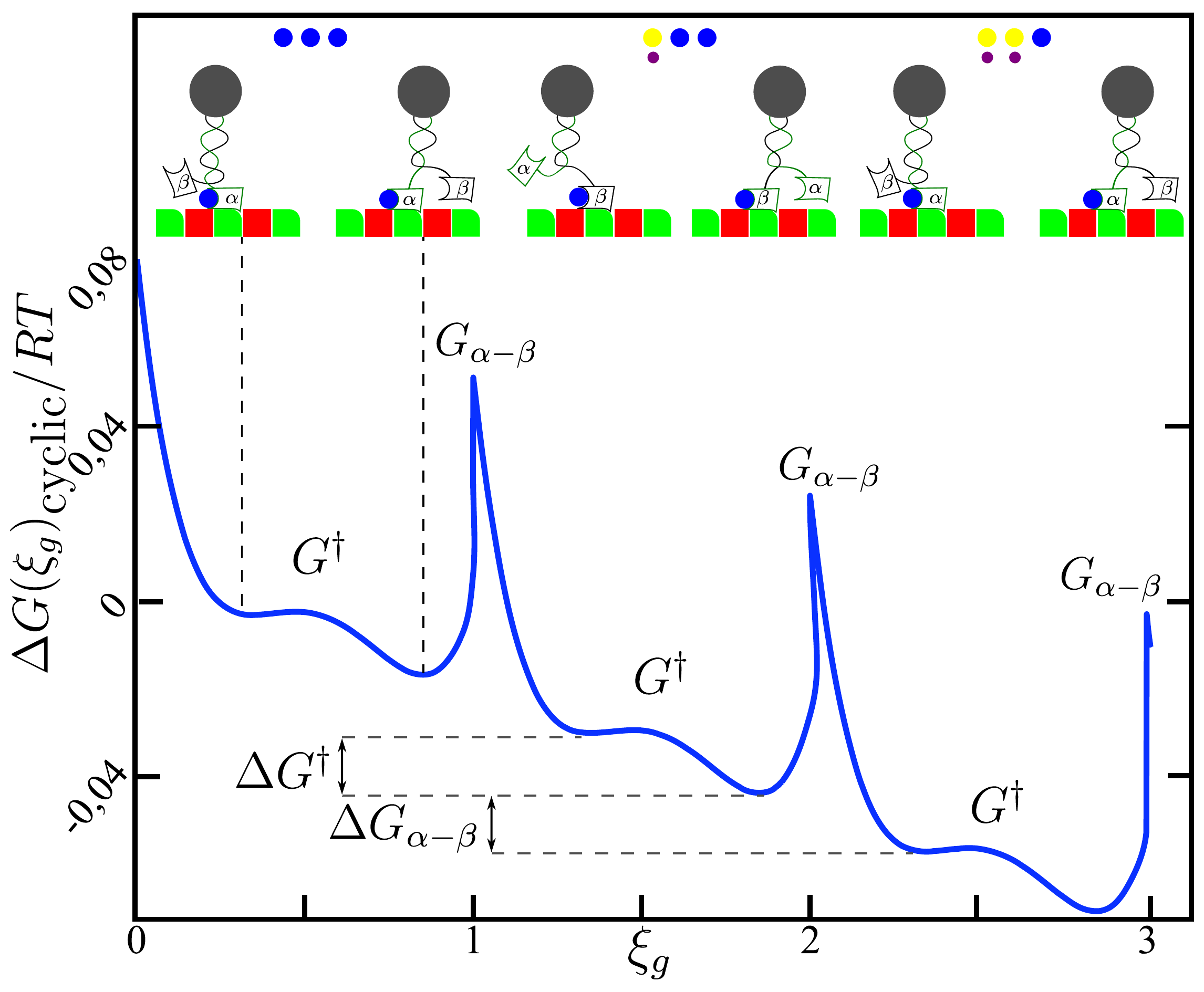}
\caption{First three steps of the free-energy landscape associated to kinesin translation as obtained from Eq. (\ref{G:4}) for the initial values: $\ n_{MKT\prime}^{o}=~0.02$,$\ n_{T}^{o}=~0.78$ and$\ n_{MK}^{o}=0.2$.  Two energy barriers appear, one related to the isomerization reaction (\ref{rec:2}), $G^\dag$, and other associated to the $\alpha-\beta$-shifting for complex $MKT^\prime_\alpha$ to $MKT_\beta$, $G_{\alpha-\beta}$ which is surmounted {\em via} ATP hydrolysis. The inset shows the free-energy differences $\Delta G^\dag$ between the states $MKT_\alpha$ and $MKT^\prime_\alpha$, and between the sates $MKT^\prime_\alpha$ and  $MKT_\beta$,  $\Delta G_{\alpha-\beta}$. The values of the other constants were taken from~\cite{Visscher2000} and~\cite{ADP} and are given in Fig. 6.}
\label{fig:3}
\end{center}
\end{figure}

Performing the Fourier series of Eq. (\ref{G:1}), the resulting expression for the catalyst energy landscape in
terms of the global reaction coordinate $\xi_g$ is
\begin{equation}\begin{aligned}\label{G:2}\displaystyle
\frac{1}{RT}\Delta G(\xi_g)_{cyclic} \simeq \frac{\mathcal A_0}{2}+\sum_{i=1}^{k}\mathcal A_n\cos(2\pi k\xi_g)\\
+\sum_{i=1}^{k}\mathcal B_n\sin(2\pi k\xi_g) -D_{eq}\xi_g+\frac{f\delta}{k_BT}\xi_g\,,
\end{aligned}\end{equation}
in which the Fourier coefficients $\mathcal A_n$ and $\mathcal B_n$ are defined by
\begin{equation}\begin{aligned}\label{An-Bn}
\mathcal A_k=\int_0^1(A \xi^4+B \xi^3+C \xi^2+D \xi)\cos(2\pi k\xi)d\xi,\\
\mathcal B_k=\int_0^1(A \xi^4+B \xi^3+C \xi^2+D \xi)\sin(2\pi k\xi)d\xi\,,
\end{aligned}\end{equation}

In order to obtain Eq. (\ref{G:2}) we have assumed that, for the first cycle, the equilibrium values of $n_i$ are equal to the initial values 
$n_i^o$, thus implying that $D=D_{eq}$. The first three steps of a kinesin motor are represented in Figure \ref{fig:3} {\em via} the Gibbs free-energy landscape 
(\ref{G:2}) of the catalytic reaction. An interesting feature of the model is that appears a second energy barrier $G_{\alpha-\beta}$ (see Figure \ref{fig:3}) that separates the state $MKT^\prime_\alpha$ from state $MKT_\beta$ [see also Eqs. (\ref{rec:3})-(\ref{rec:6})]. The amplitude of this barrier is ten times larger than that corresponding to the isomerization reaction. The fact that this second energy barrier is not too high, seems to agree with the observation \cite{Allosteric} that some kinesin motors may operate in two forms, one with directed motion by using ATP and a second non-directed motion without using ATP but using the thermal energy at their disposal.
Finally, the Gibbs free-energy difference between the states $MKT^\prime_\alpha$ and 
$MKT_\beta$ is denoted by $\Delta{G_{\alpha-\beta}}$. Numerical values for the free-energy barriers and differences were estimated using  experimental data \cite{Visscher2000,ADP} for the initial concentrations in Eq. (\ref{G:2}) and are presented in Table~\ref{tab:4}.  Since the initial values of $n_{MK}^o$, $n_T^o$ and $n_{MKT^\prime}^o$ are constant, no change of the amplitude of the barriers and energy differences appears during the catalytic process. This clearly represents an idealization that will be improved in following sections were the effect of ADP inhibition will be taken into account.
\begin{table}[htdp]
\begin{center}
\begin{tabular}{rccc}\hline\hline
Energy kJ/mol&Step 1&Step 23&Step 62\\
\hline\hline
$G^\dag$&0.012&0.012&0.012\\
$\Delta{G^\dag}$&-0.017&-0.017&-0.017\\
$G_{\alpha-\beta}$&0.083&0.083&0.083\\
$\Delta{G_{\alpha-\beta}}$&-0.005&-0.005&-0.005\\
\end{tabular}
\end{center}
\caption{Magnitudes of the energy barriers and differences associated to the free-energy landscape of the catalytic reaction as obtained from the 
model (\ref{G:2})  using the same values as those used in Figure~\ref{fig:3}.}
\label{tab:4}
\end{table}
\subsection{Motor translational velocity}\label{section:IIIB}   

It is well know that kinetics of most part of biological processes mediated by non-allosteric enzymes are well described by  Michaelis-Menten like 
equations~\cite{lehninger}. This fact may be used in the mechanism scheme (\ref{rec:1})-(\ref{rec:6}) by assuming that 
reaction (\ref{rec:3}) is the slow step in reaction-mechanism, as we did in the previous subsection. Thus, assuming that 
steps (\ref{rec:4})-(\ref{rec:6}) are fast~\cite{Visscher1999}, 
the translational velocity $v([T])$ can be obtained by multiplying the reaction velocity $w([T])$ by kinesin's step distance, $d=8\,nm$, 
and divided by the total enzyme concentration $[MK]_o$, which measures the number of active motors in a given time. 
The final expression is \cite{Visscher1999} 
\begin{equation}\label{tras-vel}\displaystyle
v([T])=\frac{dk_{cat}[T]}{K^\dag_M+[T]}\,,
\end{equation}
where $k_{cat}=k_2K^\dag/(K^\dag+1)$ is the catalyst rate constant, denoting the maximum number of enzymatic reactions catalyzed per second~\cite{Visscher1999}. 
$K^\dag_M$  is the pseudo-Michaelis-Menten constant~\cite{Visscher2000} 
defined by $K^\dag_M=k_1K^\dag/(k_{-1}+k_2K^\dag)$.  
A more detailed derivation of Eq. (\ref{tras-vel}) is reported in \cite{Visscher1999}.
Experimental results have shown that the Michaelis-Menten approach is sufficiently accurate for
large concentration values of ATP \cite{Visscher2000,ADP}. 
\section{Quantitative influence of ADP on processivity}\label{sec:IV}

Experimental  studies have reported that the presence of ADP and $P_i$ reduces the 
processivity of kinesins by inhibiting the formation of the enzyme-substrate complexes $MKT$ and 
$MKT^\prime$, which are essential in turn to the motility of the motor \cite{ADP,Yajima,Romberg,premature}. 
On the basis of these observations, the previous analysis can be modified appropriately in order to account 
for this finite processivity of kinesins, that is, their limited number of cycles. 

Several types of ADP and $P_i$ inhibitions are possible in principle, like competitive or non-competitive 
depending on their release order in the corresponding kinesin head \cite{ADP}. Each type of inhibition will modify 
in different way the expression for the kinesin velocity. The explicit form in which these corrections will
appear are described in detail in Appendix D. However, it was experimentally found in Ref.  \cite{ADP} 
that the effect of the non-competitive and competitive inhibitions by $P_i$ are only relevant at low ATP 
concentrations, which suggests that they could be observable only for a short time period at the end of the process. This means that competitive inhibition by ADP can be considered as the leading inhibition effect of the process, 
as it will be considered here. The generalization of the model to include the other type of inhibition mechanisms
is straightforward.

In order to deduce the free-energy landscape containing the effect of ADP inhibition, it is necessary to first analyze how motor displacement
velocity changes, and how the concentrations of ATP and ADP evolve in time. 
\subsection{Motor translational velocity in the presence of ADP inhibition}\label{sec:IVA}

In accordance with the previous considerations, we will assume that ADP inhibition can be described as a competitive reaction since
kinesin active sites seem to be affine to ATP or ADP in an exclusive way. This hypothesis modifies the reaction scheme  (\ref{rec:1})-(\ref{rec:6})
by including the parallel reaction
\setcounter{equation}{0}
\begin{subequations}\label{rec:1.1}
\renewcommand{\theequation}{\theparentequation-\arabic{equation}}
\begin{equation}\displaystyle
MK_\alpha^{}+D\ce{<=>[K_I]}MKD_\alpha^{}\,.
\end{equation}
\end{subequations}
\setcounter{equation}{19}
\begin{figure}[!htbp]
\begin{center}
\includegraphics[width=240pt]{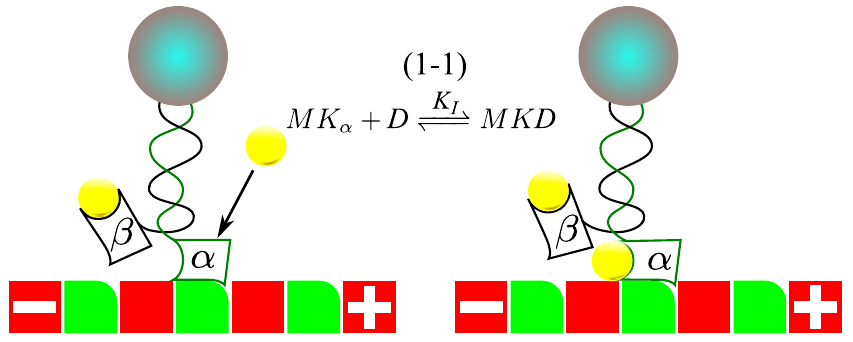}
\caption{Schematic representation of parallel inhibition reaction, Eq. (1-1).}
\label{fig:7}
\end{center}
\end{figure}
Following the same procedure as in section 3.2, it may be shown that the displacement velocity of the kinesin in the presence of inhibition is
\begin{equation}\label{eq:16}\displaystyle
v([T],[D])=\frac{dk_{cat}[T]}{K^\dag_M(1+[D]/K_I)+[T]}\,,
\end{equation}
where $K_I$ is the inhibition constant~\cite{ADP}.  It is important to emphasize that this relation assumes step (\ref{rec:3}) as the slow step that determines the behavior of the kinetics, but it does not takes into account that ADP is a product of the whole reaction scheme. As a consequence of this, Eq. (\ref{eq:16}) can not be used to evaluate the ADP production kinetics, this is more probably determined by the last step (\ref{rec:6}).

In order to reconstruct the free energy landscape of the catalytic process in presence of inhibition by ADP, it is necessary to implement the time dependence of ATP and ADP concentrations since in the overall process the initial concentrations after each cycle ($n_T^0$, $n_{MK}^0$ and $n_{MKT}^0$) change as a function of time. 

For this purpose, it is necessary to solve the reaction velocity equation $w=-d[T]/dt$. The solution is given in implicit form by the expression
\begin{equation}\label{[T]t}\displaystyle
K^{\dag} \ln\left|\frac{[T]_t}{[T]_o}\right|+[T]_t-[T]_o=-v_{max}t,
\end{equation}
where the subindex $t$ indicates time dependence, and $v_{max} \equiv k_{cat}[MK]_o$. From  these relation it follows that $[T]_t$ is given 
by the  $W$ Lambert function \cite{Lambert}. 

For the values of the parameters previously used, it may be shown by directly evaluating the function (or by numerically solving the equation) that Eq. (\ref{[T]t}) can be well approximated by
\begin{equation}\label{T-de-t}\displaystyle
[T]_t \simeq [T]_o-v_{max}t.
\end{equation}

Here, it is convenient to mention that  due to the crowded nature of the intracellular 
medium, transport of ATP and ADP by diffusion becomes a very slow and probably even a confined process, 
as it has been observed experimentally for chromosomes \cite{marshall} and theoretically explained 
in \cite{JCPfinite}. As a consequence of this the theoretical treatment of the process can
be performed, in a first approximation, without taking into account spatial heterogeneity or ATP production.

\begin{figure}[!ht]
\begin{center}
\includegraphics[width=240pt]{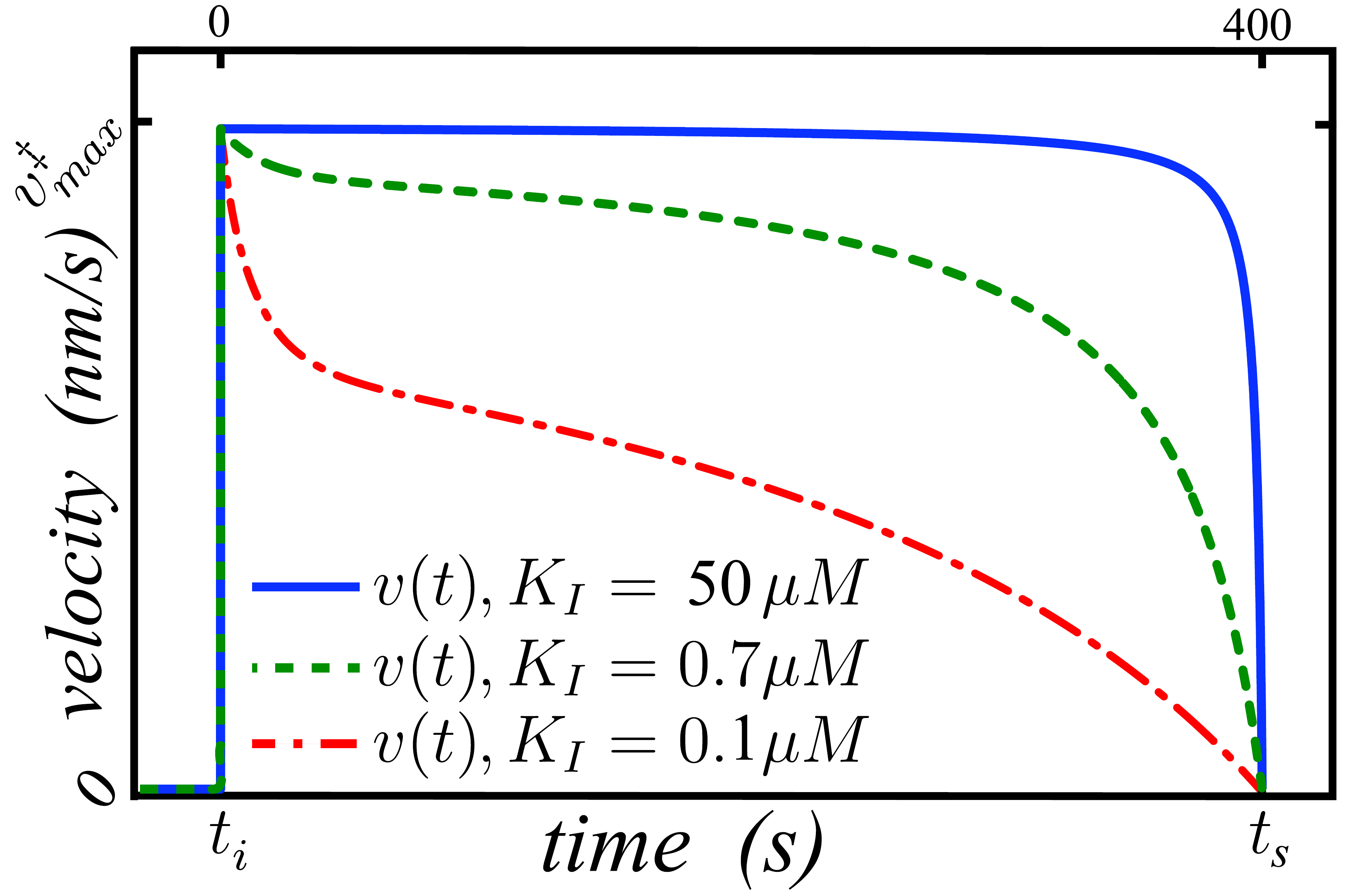}
\caption{Translational velocity of a kinesin motor as a function of time given by Eq. (\ref{tras-vel-2}) with the following values for the constants entering into Eqs. (\ref{tras-vel-2}) and (\ref{eq:21}):  $\ [T]_o=~1\,mM$,$\ [D]_o=~0$,$\ v^{\ddag}_{max}=900\,nm\,s^{-1}$, $v_{max}=11.25\mu{M}\,s^{-1}$, $[MK]_o=100nM$, $K_{M}^\dag=28\mu{M}$, $\ k_{cat}=113\,s^{-1}$, $k_1=4.0\mu{M}^{-1}s^{-1}$, $k_{-1}=3.8\times10^{-5}s^{-1}$, $\ k_5=5.6\times10^{3}\,s^{-1}$, $K_o=0.0017$ were taken also from Refs.~\cite{Visscher2000,ADP}. The value $\ K_M^{\ddag}=2.24\,\mu{m}$ was calculated using its definition  $K^\ddag_M=dK^\dag_M/[MK]_o$.}
\label{fig:8}
\end{center}
\end{figure}

On the other hand, according to experimental results \cite{ADP-production-1,ADP-production-2} and the previous considerations on Eq. (\ref{rec:6}), it follows that the time behavior of the ADP concentration can be modeled by the first order kinetics of the form
\begin{equation}\label{D-de-t}\displaystyle
[D]_t=[D]_o+[T]_o(1-e^{-k_5t}).
\end{equation}

Using equations (\ref{T-de-t}) and (\ref{D-de-t}) we obtain the following expression for the average displacement velocity in terms of time
\begin{equation}\label{tras-vel-2}\displaystyle
v(t)=\frac{v_{max}^{\ddag}(t_s-t)}{h_t+(t_s-t)}\,,
\end{equation}
where we have defined the maximum displacement velocity  $v^\ddag_{max} \equiv dk_{cat}=dv_{max}/[MK]_o$, the
stopping time $t_s=[T]_o/v_{max}$ and the function
\begin{equation}\label{eq:21}\displaystyle
h_t=\frac{\left[K^\ddag_M(1+ [D]_t/K_I)\right]}{v_{max}^{\ddag}}\,,
\end{equation}
where  $K^\ddag_M=dK^\dag_M/[MK]_o$ and $ [D]_t$ is given in (\ref{D-de-t}). Eq. (\ref{tras-vel-2}) is shown in Fig.~\ref{fig:8} for different values of the 
inhibition constant $K_I$. It is notable that when inhibition is small ($K_I$ large) the displacement velocity is almost constant during the elapsed 
time of the process and near to the saturation velocity. When inhibition increases ($K_I$ decreases) the velocity departs from saturation levels and 
becomes non constant. These result may be important for biological processes such as exocytosis \cite{Exocytosis}. In Fig.~\ref{fig:8} we have 
assumed that the process starts instantaneously when there exist saturation levels of ATP and $\ce{Ca+2}$ signalment have occurred \cite{steinhardt}. 
We have modeled this situation by means of  a Heaviside function at a initial time $t_i$ in Eq. (\ref{tras-vel-2}): $v(t)=\Theta(t-t_i)v(t)$.

We can estimate the time duration of a given process using  Eq. (\ref{tras-vel-2})  with the values used on Fig.~\ref{fig:8}. We obtain $t_s=[T]_o/v_{max}=400s$. 
Because the stopping time has been calculated by assuming that the translational velocity depends on the total concentration of the enzyme $MK$, it contains
the effect of motor cooperativity and has to be compared with cellular processes that probably involve the participation of several motors. This may be
the case of some exocytosis-endocytosis process. In fact, the stopping time we obtained agrees well with the characteristic times observed in experiments 
studying membrane resealing, where the active transport of vesicles from the inner part of the cell to the plasma membrane is very probable (see Ref. \cite{Exocytosis}). 
Another important result emerging from the present analysis is related to the fact that the constant $K^\ddag_M \equiv d\,K_M/[MK]_0$ can be interpreted 
as the average distance of advance until its velocity down to half of saturation velocity $v^\ddag_{max}$. The average travelled distance corresponding to 
the estimated stopping time is $K^\ddag_M  \simeq 2.4 \mu\,m$, thus implying the consumption of about $\sim300$ ATP molecules per motor. These results 
also compare well with experimental evidence \cite{hanckock}. 

\subsection{Single motor free-energy landscape in the presence of ADP inhibition}\label{sec:IVB}

The analysis and results obtained in previous sections bring us key facts to incorporate ATP consumption and ADP production into the free-energy model. 
As we showed, the increase of ADP concentration may modify the equilibrium on (\ref{rec:2}) because a parallel inhibition reaction between the $MK$ complex 
and ADP is also present~\cite{ADP}. However, one may still assume that free enzyme is in a steady state in similar way to the Michaelis-Menten 
kinetic~\cite{Visscher1999,ADP,Helfferich,SRlogan,nature-Inhibition}. 
\begin{table}[!htb]
\begin{center}
\begin{tabular}{lccc}
\hline
\multicolumn{4}{l}{Time \hfill Molar fractions\hfill \ \ }\\[1pt]
\hline\hline\\[1pt]
$\ t$&$\ n_{MK}^o$&$\ n_{T}^o(t)$&$\ n_{MKT^\prime}^o(t)$\\[3mm]
$\ t+\tau$&$\ n_{MK}^o-\xi$&$\ n_{T}^o(t+\tau)-\xi$&$\ n_{MKT^\prime}^o(t+\tau)-\xi$
\end{tabular}
\caption{Time dependent stoichiometric ratios. The molar fraction of $MK$ is independent of  time since it is considered the free enzyme.}
\label{tab:2}
\end{center}
\end{table}

The procedure to obtain the single motor catalytic energy landscape in the presence of ADP as an inhibitor is similar to the one followed in Section 3. 
However, in the present case we have to consider that after each single cycle, the initial values for the mass fractions $n_{MK}^o$, $n_T^o$ and 
$n_{MKT^\prime}^o$ have to be recalculated by taking the corresponding values $n_{MK}^o(\tau)$, $n_T^o(\tau)$ and $n_{MKT^\prime}^o(\tau)$. 
Thus, for any other subsequent cycle, the data of Table~\ref{tab:1}  become: 
$n_i^o(t)$ and $n_i^o(t+\tau)-\xi$, see Table~\ref{tab:2}. 

This time dependence enters because the initial concentration of ATP decreases with time $\tau$ as the catalytic reaction takes place and produces ADP, Eqs. (\ref{T-de-t}) and (\ref{D-de-t}).  Hence, the coefficients $\mathcal A_n$, $\mathcal B_n$ in (\ref{G:2}) become functions of time $\tau$ through the initial concentrations after each cycle, $n_i^o(\tau)$. This may be 
represented schematically by the subindex $\tau$ in coefficients:  $B_\tau$, $C_\tau$ and $D_\tau$ (see Appendix B for details).

The explicit form of this time dependence can be derived following Ref. \cite{Transition-time}. It may be shown (see Appendix B)  that ATP 
concentration decreases following the relation:  $n^o_{T}(\xi_g)=n^o_{T}-k_{cat}\tau\xi_g$, which can be obtained from Eq. (\ref{T-de-t}) and considering 
the existing bijection between $\tau$ and $\xi_g$ indicated in Figure \ref{fig:9}.
In addition to this, the parallel inhibition reaction between the $MK$ complex and ADP (ADP competes for the same active site of kinesin as ATP) 
avoids the formation of both enzyme-substrate complexes $MKT$ and $MKT^\prime$. Thus, this inhibition increases with $\xi_g$ because the 
whole reaction cycle produces ADP and therefore reduces $MKT^\prime$. This decrease of  $MKT^\prime$ cannot be directly inferred 
from the kinetic equation since the reaction velocity do not depends on the mass fraction of the complex $MKT^\prime$. However, in similar way as 
estimated for the ATP decrease, we can assume that the initial mass fraction $n^o_{MKT^\prime}$ decreases in linear form with  $\xi_g$, that is: 
$n^o_{MKT^\prime}(\xi_g)=n^o_{MKT^\prime}-K_o\xi_g$, where the constant $K_o$ is the equilibrium constant without load [see Eq.~(\ref{rec:2})].

Thus, after performing the Fourier series having in mind the previous considerations, the resulting expression for the catalyst energy landscape is
\begin{equation}
\begin{aligned}\label{G:3}
\frac{1}{RT}\Delta G(\xi_g,\tau)_{cyclic} \simeq \frac{\mathcal A_0(\tau)}{2}+\sum_{i=1}^{k}\mathcal A_k(\tau)\cos(2\pi k\xi_g)\\
+\sum_{i=1}^{k}\mathcal B_k(\tau)\sin(2\pi k\xi_g) -D_{eq}\xi_g+\frac{f\delta}{k_BT}\xi_g\,,
\end{aligned}
\end{equation}
in which the Fourier time dependent coefficients $\mathcal A_n(\tau)$ and $\mathcal B_n(\tau)$ are defined by
\begin{equation}\begin{aligned}\label{An-Bn-t}
\mathcal A_k(\tau)=\int_0^1(A\xi^4+B_\tau\xi^3+C_\tau\xi^2+D_\tau\xi)\cos(2\pi k\xi)d\xi,\\
\mathcal B_k(\tau)=\int_0^1(A\xi^4+B_\tau \xi^3+C_\tau\xi^2+D_\tau\xi)\sin(2\pi k\xi)d\xi\,.
\end{aligned}\end{equation}

Establishing now the bijection between $\tau$ and $\xi_g$ according to Figure~\ref{fig:9}, we may 
recast Eq. (\ref{G:3}) in its final form
\begin{equation}
\begin{aligned}\label{G:4}\displaystyle
\frac{1}{RT}G(\xi_g)_{cyclic} \simeq \frac{\mathcal A_0(\xi_g)}{2} +\sum_{i=1}^{k}\mathcal A_k(\xi_g)\cos(2\pi k\xi_g)\\
+\sum_{i=1}^{k}\mathcal B_k(\xi_g)\sin(2\pi k\xi_g)-D_{eq}\xi_g+\frac{f\delta}{k_BT}\xi_g.
\end{aligned}
\end{equation}
As in Section 3, in order to obtain Eq. (\ref{G:4}) we have assumed that, for the first cycle, the equilibrium values of 
$n_i$'s are equal to the initial values $n_i^o$'s, thus implying $D=D_{eq}$. Equation (\ref{G:4}) is very general since models 
both, every single step during the translation of a single kinesin motor, but also the overall process and even the finite 
processivity of the motor, as we will explain in the next. 
\begin{figure}[!htb]
\begin{center}
\includegraphics[width=240pt]{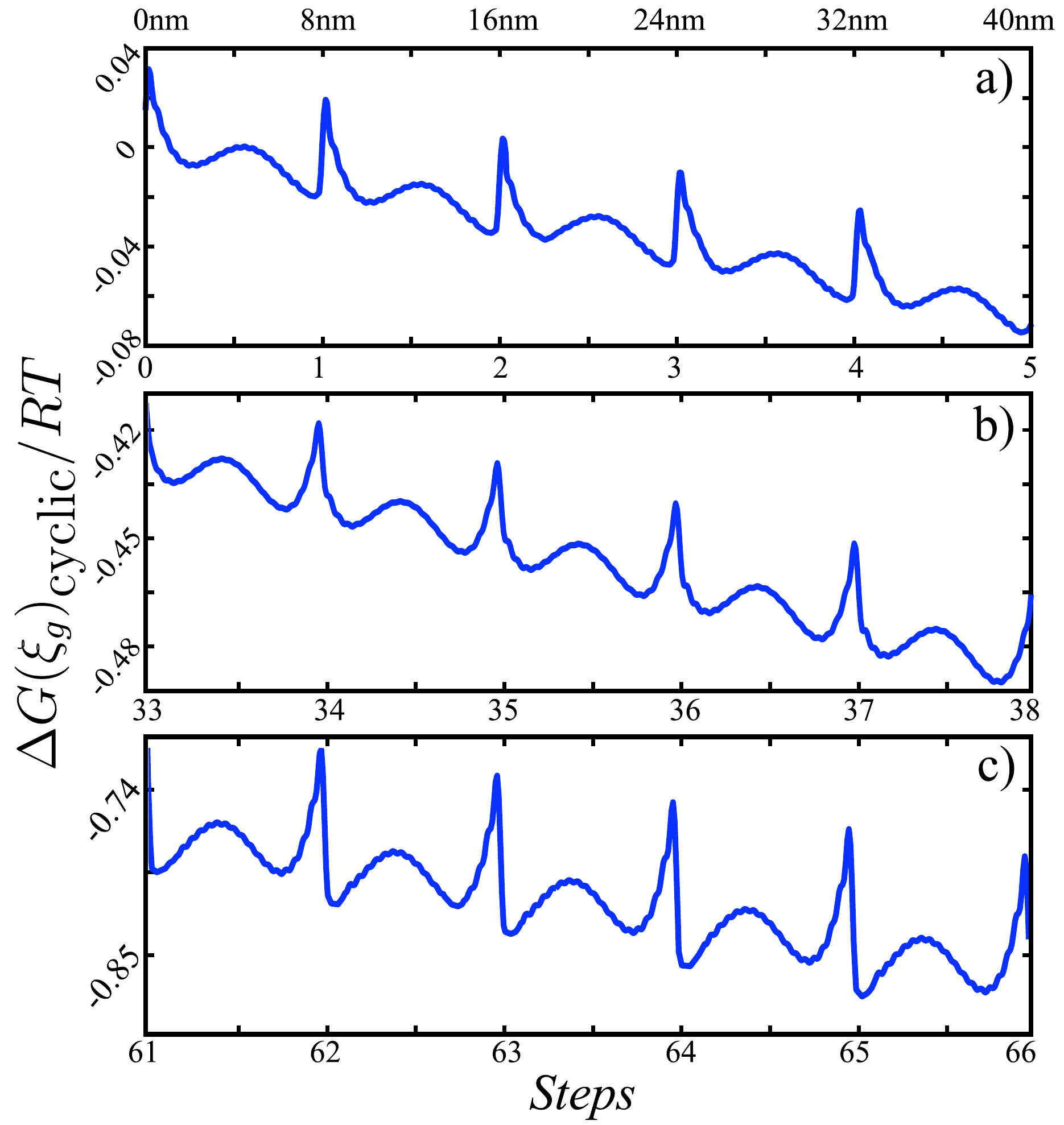}
\caption{Free energy landscape obtained \emph{via} the Fourier expansion Eq. (\ref{G:4}) with $k=25$ modes for a catalytic reaction of many cycles with no cargo. The equilibrium values of the mass fractions used were: $\ n_{MKT^\prime}^{eq}=~0.02$,$\ n_{T}^{eq}=~0.78$ and$\ n_{MK}^{eq}=0.2$. a) from cycle 1 to 5 and corresponding $\ nm$ displacement. b) step 33 to 38. c) step 61 to 66. The values of the constants used are given in the caption of Figure 6.}
\label{fig:10}
\end{center}
\end{figure}
\section{Discussion}\label{sec:V}

\begin{figure}[!htb]
\begin{center}
\includegraphics[width=240pt]{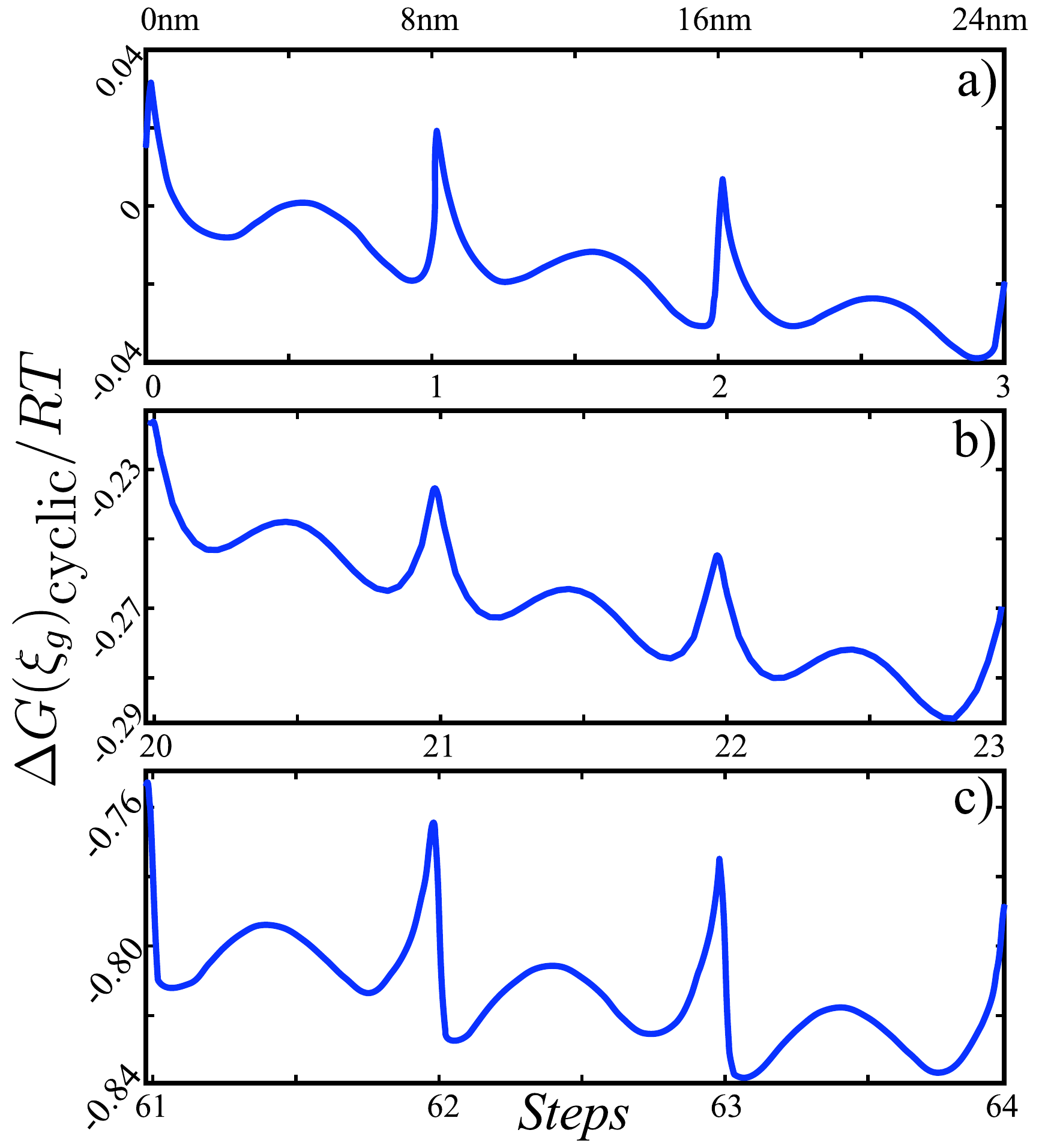}
\caption{Free energy landscape obtained \emph{via} the Fourier expansion Eq. (\ref{G:4}) with $k=25$ modes for a catalytic reaction of many cycles with cargo $f=0.02$pN. The equilibrium values of the mass fractions were $\ n_{MKT^\prime}^{eq}=~0.02$,$\ n_{T}^{eq}=~0.78$ and $\ n_{MK}^{eq}=0.2$. a) from cycle 1 to 3 and corresponding $\ nm$ displacement. b) step 20 to 23. c) step 61 to 64. The values of the constants used are given  in the caption of Figure 6.}
\label{fig:11}
\end{center}
\end{figure}

Three different sequences of steps of the Gibbs free-energy landscape (\ref{G:4}) of the catalytic reaction associated to kinesin translation are 
represented in Figures \ref{fig:10} and \ref{fig:11} for the free of cargo and cargo cases, respectively.
In similar way as in Section 3, Eq. (\ref{G:2}), in the present case we find two energy barriers associated to the translations of the kinesin. The first
barrier separates the isomeric states $MKT$ and $MKT^\prime$, whereas the second one separates the $\alpha -\beta$-shifted states.
The important fact to emphasize here is that the free-energy landscape (\ref{G:4}) contains amplitude coefficients $\mathcal{A}_n$ and 
$\mathcal{B}_n$ that depend explicitly on 
the global degree of reaction $\xi_g$ and the initial mass fractions $n^0_i$. Since these dependences take into account the feedback inhibition 
by the product of the reaction ADP, which increases with $\xi_g$, these amplitudes of cosine and sinuous terms of the expansion increase after many steps. 
\begin{table}[htdp]
\begin{center}
\begin{tabular}{rccc}\hline\hline
Energy kJ/mol&Step 1&Step 23&Step 62\\
\hline\hline
$G^\dag$&0.012&0.018&0.023\\
$\Delta{G^\dag}$&-0.017&-0.014&0.009\\
$G_{\alpha-\beta}$&0.083&0.088&0.135\\
$\Delta{G_{\alpha-\beta}}$&-0.005&-0.07&-0.02\\
\end{tabular}
\end{center}
\caption{Comparison of energy requirements and energy differences in three different steps of the catalytic reaction
as predicted by Eq. (\ref{G:4}). These energies were obtained for the same values as those used in Fig.~\ref{fig:10}.}
\label{tab:5}
\end{table}
As a consequence of this effect, the magnitude of both free energy barriers increases and also changes the tilting of the overall potential. 

These two features of  (\ref{G:4}) have the consequence that, after a finite number of steps, the displacement of the motor stops. Specifically, this occurs because  the fast equilibrium between $MKT$ and $MKT^\prime$ states is inverted in the sense that the {\em backward} state $MKT$ more favorable than the {\em forward} state $MKT^\prime$, see the last steps in Figures \ref{fig:10}c and \ref{fig:11}c.  In accordance with the energy minimizing statement, this change of relative values of the corresponding free-energies implies that the motors cannot follow in its motion. In addition, the amplitude of both barriers increases as the reaction takes place. This fact slows down the activated process between $MKT$ and $MKT^\prime$ sates and finally makes more improbable  that the energy coming from ATP hydrolysis is enough to overcome the barrier of the $\alpha -\beta$-shifted states, see the difference between Figures \ref{fig:10}a and \ref{fig:10}c, and of Figures \ref{fig:11}a and \ref{fig:11}c. Using the data of the previous analysis, we predict that the free of cargo motion is characterized by a processivity of about 66 steps, which corresponds to an energy consumption of $66$ ATP molecules and a traveled distance of $528\,nm$. This number suggests that a motor traveling $2.4 \mu m$ may stop about for times. In the case when a $0.02\,pN$ cargo is applied the processivity reduces to $64$ steps. Finally, the load term $f\delta/k_BT$ can be viewed as the ratio between the work done  by the kinesin to move such cargo and the thermal energy available from surroundings (intracellular space), along the overall reaction. Since the load  term is positive and the linear equilibrium term $D_{eq}$ is negative, we can conclude that the presence of a cargo reduces the processivity of motor by  stopping the reaction before with respect to the case without load.  

Numerical values for the free-energy barriers were estimated with Eq. (\ref{G:4}) and are presented in Table~\ref{tab:5}, where we quantify the 
energy consumption of kinesin's translational motion for different times (steps). The change of sign of $\Delta G^\dag$ (at step 62) shows that
the state $MKT^\prime$ becomes less favorable than $MKT$. This effect is also present with load (Fig.~\ref{fig:11}) at an earlier step.
The increase of the energy barriers $G^\dag$ and $G_{\alpha-\beta}$ is clear.
\section{Conclusions}\label{sec:VI}

In this work we proposed two biochemical models providing the kinetic and energetic explanations of the processivity 
dynamics of kinesin, myosin and dinein-type molecular motors. Our approach is based on a well known model 
describing kinesin dynamics and considers the presence of a competitive inhibition reaction by ADP. 
The two models discussed constitute a powerful tool to understand and describe in quantitative form the dynamics of translational 
molecular motors and therefore may be useful to give more precise descriptions of some cellular processes mediated
by kinesins or myosins such as, for instance, exocytosis and endocytosis. 

We provided a new analytical procedure to reconstruct a continuous free-energy landscape of the cycle catalyst process starting
from the corresponding biochemical reaction model. The obtained  free-energy landscape is valid for a single motor and allows one 
to predict the heigh of the main free-energy barriers associated to the motion and the total number of steps given
by the molecular motor for given physical and chemical conditions of the surroundings. That is the processivity of the motor
is explained by means of energetic considerations.

In addition, a collective description of the dynamics of translational molecular motors is also provided that allows to determine 
an analytical expression for the associated translational velocity as well as an expression for the stopping time of the molecular motors 
in terms of time and ATP concentration.

Motor's processivity is discussed in quantitative form by using experimental data. The number of
steps we predict (60-66) agrees well with experimental observations and provides important quantitative information of the 
energy consumption during the process.  In similar way, the average velocity and the time duration of a collective process 
is estimated via the kinetic description and also agrees ($t_s \simeq 400s$) with experimental reports on 
secretion processes mediated by kinesins and myosins.  In this respect, the time dependence of motors' velocity may be
very important since it may be used in coarse-grained models describing intracellular transport \cite{Rubi1}.
\begin{acknowledgments}
We acknowledge Profs. A. P\'erez-Madrid and K. Michaelian by critically reading this manuscript.  We also thank UNAM-DGAPA for partial financial support of Grant No. ID100112-2 and CONACYT.
\end{acknowledgments}
\appendix
\section{Derivation of equations (\ref{dG:2})-(\ref{dG:3})}\label{app:A}

The main supposition in order to derive Eqs. (\ref{dG:2}) and (\ref{dG:3}), is to assume that the evolution of the coupled reactions 
(\ref{rec:1}) and (\ref{rec:2}) can be described by using a single degree of reaction $\xi$. This assumption is plausible because 
reaction (\ref{rec:2}) is in equilibrium, and therefore it allows to relate $MKT$ with $MKT^\prime$ concentrations through the reaction 
constant $K^\dag$. The main consequence of this assumption is that a fourth order polynomial in $\xi$ is recovered for the 
Gibbs free-energy associated to the enzyme reaction scheme.

Let us start by expanding the free-energy change of the reaction (\ref{rec:1}) and using that elementary reactions obey
the following relation between the chemical affinity $A_j$ of $j$-th reaction and the corresponding degree of reaction:  $dG_j=-A_jd\xi_j$.
Then, it follows that for the reaction (\ref{rec:1}) the change on Gibbs free-energy is 
\begin{equation}\begin{aligned}\label{eq:A1}\displaystyle
dG_1=-\left(\nu_{MK_1}\mu_{MK_1}+\nu_{T_1}\mu_{T_1}\right.+\cdots\\
+\left.\nu_{MKT_1}\mu_{MKT_1}\right)d\xi_1\,,
\end{aligned}\end{equation}
where the stoichiometric coefficients of reaction (\ref{rec:1}) are:  $\nu_{MK_1}=1$, $\nu_{T_1}=1$ and $\nu_{MKT_1}=-1$. 
Using now an ideal approximation for all chemical potentials ($\mu = k_BT \ln|n|$) we get
\begin{equation}\label{eq:A2}\displaystyle
dG_1=-RT\ln \left|\frac{n_{MK}^{\nu_{MK_1}}}{n_{MK_{eq}}^{\nu_{MK_1}}}\cdot\frac{n_{T}^{\nu_{T_1}}}{n_{T_{eq}}^{\nu_{T_1}}}\cdot\frac{n_{MKT}^{\nu_{MKT_1}}}{n_{MKT_{eq}}^{\nu_{MKT_1}}}\right|d\xi_1\,.
\end{equation}

Now, since we have assumed that reactions (1) and (2) are coupled, then it follows that at every time the concentration of the $MKT$ complex can be 
calculated as: $n_{MKT}=n_{MKT}^o+\xi_1$, or $n_{MKT}=n_{MKT}^o-\xi_2$, with $\xi_2$ the degree of reaction of the isomerization reaction. 
Comparing these two expressions we conclude that
\begin{equation}\label{eq:A3}\displaystyle
\xi_1=-\xi_2\,.
\end{equation}
The above relation is fundamental in the calculation of the free-energy Eq. (\ref{dG:3}) since it demonstrates that both reactions 
can occur with a same degree of reaction $\xi$.

Now if we use equilibrium approximation for Eq. (\ref{rec:2}), we know that the equilibrium constant is defined as \cite{Prigogine} 
\begin{equation}\label{eq:A4}\displaystyle
K_{eq}=\prod \frac{n_i^{|\nu_i^*|}}{n_i^{\nu_i}}\,,
\end{equation}
where $\nu_i$ denotes the stoichiometric coefficient for reactants and $\nu_i^*$ the stoichiometric coefficient for products. For simplicity, in the
following we will consider that all $\nu^*$ coefficients are negative. Hence, using this definition in reaction (\ref{rec:2}) we get 
\begin{equation}\label{eq:A5}\displaystyle
n_{MKT}^{\nu_{MKT_2}}=\frac{n_{MKT^\prime}^{|\nu_{MKT^\prime_2}|}}{K^\dag}\,,
\end{equation}
where the $MKT$-complex acts as the reactant and the complex $MKT^\prime$ as the product: $\nu_{MKT_2}=1$ and $\nu_{MKT^\prime_2}=-1$. 
Noticing that for the reaction (1): $\nu_{MKT_1}=-1$, whereas that for the reaction (2): $\nu_{MKT_2}=1$, then by using the result of (\ref{eq:A3}) we can 
establish the following relation 
\begin{equation}\label{eq:A6}\displaystyle
(n_{MKT}^o+\xi_1)^{|\nu_{MKT_1}|}=(n_{MKT}^o-\xi_2)^{\nu_{MKT_2}}\,,
\end{equation}
and, with Eq. (\ref{eq:A5}) and (\ref{eq:A6}) we can also establish
\begin{equation}\label{eq:A7}\displaystyle
(n_{MKT^\prime}^o+\xi_2)^{|\nu_{MKT^\prime_2}|}=(n_{MKT^\prime}^o-\xi_1)^{\nu_{MKT^\prime}}\,,
\end{equation}
where $\nu_{MKT^\prime}=1$. Finally, from (\ref{eq:A5})-(\ref{eq:A7}) it follows that
\begin{equation}\label{eq:A8}\displaystyle
n_{MKT}^{\nu_{MKT_1}} = (n_{MKT}^o+\xi_1)^{\nu_{MKT_1}}=\frac{(n_{MKT\prime}^o-\xi_1)^{\nu_{MKT\prime}}}{K^\dag}\,.
\end{equation}
Eq. (\ref{eq:A8}) can be substituted into Eq. (\ref{dG:2}) in order to obtain Eq. (\ref{dG:3}), that is the effective Gibbs free-energy change for the coupled 
reactions (1) and (2) expressed in terms of the activated complex $MKT^\prime$, that is the measurable quantity \cite{ADP}. 
\section{Rescaling the initial mass fractions after each cycle}\label{app:C}

During the advance of the reaction along the global reaction coordinate ADP is produced and ATP concentration decreases by 
following the relation  $n^o_{T}(\xi_g)=n^o_{T}-k_{cat}\tau\xi_g$, which follows directly from Eq. (\ref{T-de-t}).
Since there is a parallel inhibition reaction between the $MK$ complex and ADP (ADP competes for the same active site of ATP), 
this reaction do not allows the formation of both $MKT$ and $MKT^\prime$ complexes. The effect of inhibition increases with $\xi_g$ 
since the system is producing ADP. This decreasing behavior cannot be directly inferred from a kinetic equation, since the corresponding
reaction velocity does not depends on the concentration of the complex $MKT^\prime$. However we can make the assumption that the initial mass 
fraction $n^o_{MKT\prime}$ decreases in linear form with $\xi_g$, that is, $n^o_{MKT\prime}(\xi_g)=n^o_{MKT\prime}-K_o\xi_g$. This dependence
is similar to that of $n^o_{T}(\xi_g)$, that follows from transition time theory~\cite{Transition-time}. Taking into account these considerations, 
we see that the Fourier coefficients depends on $\xi_g$ in such a way that Eq. (\ref{dG:3}) transforms into
\begin{equation}\begin{aligned}\label{dG-inh}\displaystyle
\frac{dG}{RT}=-\left\{[n^o_{MK}-\xi][n^o_{T}(\xi_g)-\xi][n^o_{MKT\prime}(\xi_g)-\xi]\right.\\
\left.-[n^{eq}_{MK}n^{eq}_{T}n^{eq}_{MKT\prime}]+f\delta/k_BT\right\}d\xi\,.
\end{aligned}\end{equation}
Starting from Eq. (\ref{dG-inh}), we perform an integration over $\xi$ as in Section 3. The Fourier series can also be performed in which the 
Fourier coefficients are now given by the relations
\begin{equation}\label{Anx}\displaystyle
\mathcal A_n(\xi_g)=\!\!\!\int_0^1\!\!\!\!\cos(2\pi{n}\xi)\!\!\int_0^\xi\!\!\!(n-\chi)(n-\xi_g-\chi)(n-\xi_g-\chi)d\chi d\xi,
\end{equation}
and
\begin{equation}\label{Bnx}\displaystyle
\mathcal B_n(\xi_g)=\!\!\!\int_0^1\!\!\!\!\sin(2\pi{n}\xi)\!\!\int_0^\xi\!\!\!(n-\chi)(n-\xi_g-\chi)(n-\xi_g-\chi)d\chi d\xi.
\end{equation}
From Eqs. (\ref{Anx}) and (\ref{Bnx}) it follows that, after performing integrations, linear and quadratic terms in $\xi_g$ will remain.
Thus, $\mathcal A_n(\xi_g)$ and $\mathcal B_n(\xi_g)$ will be second order polynomials in $\xi_g$, and not constants as in  Eq. (\ref{G:2}). 
For notation's simplicity,we dropped out explicit dependencies on  $k_{cat}$ and $K_o$ in the last equations.
\section{ Effects of non competitive inhibition by ADP and inhibition by orthophosphate.}\label{app:C}

As mentioned in the text, one might also consider non-competitive inhibition by ADP, 
competitive and non-competitive inhibitions by orthophosphate and also a mix ADP$\circ P_i$ non-competitive and 
competitive inhibitions. These effects will introduce new inhibition constants: $K_{II}^{^{ADP}}$ for non-competitive 
ADP inhibition,  $K_{I}^{^{P}}$ for competitive $P_i$ inhibition, $K_{II}^{^{P}}$ for non-competitive $P_i$ inhibition~
\cite{ADP} and $K_{II}^{^{ADP\circ P}}$ and $K_{I}^{^{ADP\circ P}}$ for the mixed cases. Considering
these effects the numerator and denominator of Eq. (\ref{eq:16}) become
\begin{equation}
dk_{cat}\left\{1+\frac{[D]}{K_{II}^{^{ADP}}}+\frac{[P]}{K_{II}^{^{P}}}+\frac{[D][P]}{K_{II}^{^{ADP\circ P}}}\right\}^{-1}\,
\end{equation}
and
\begin{equation}
K_M^\dag\left\{1+\frac{[D]}{K_{I}}+\frac{[P]}{K_{I}^{^{P}}}+\frac{[D][P]}{K_{I}^{^{ADP\circ P}}} \right\}\,,
\end{equation}
respectively. The above expressions together with the time dependence of the concentration of orthophosphate 
$[P]_t$ leads to a new expression for $v_{max}^{\ddag}$ and $h_t$ in Eq.~(\ref{tras-vel-2}). However,
as it was  previously indicated,  the effect of these type of inhibitions is significative only at low ATP 
concentrations, that is, only in a time interval near the end of the process~\cite{ADP}. 
\section*{References}

\end{document}